\documentclass[aps,prl,reprint,showpacs,preprintnumbers,amsmath,amssymb]{revtex4-1}

\usepackage{graphicx}
\usepackage{dcolumn}
\usepackage{bm}
\usepackage{color}

\usepackage{amsmath}
\usepackage{amsfonts}
\usepackage{mathtools}
\allowdisplaybreaks

\begin{document}
\def\v#1{{\bf #1}}

\title{Competing Superconducting States in Strong Ferromagnets}

\author{Marios Georgiou}\email{Electronic address: mgeor@mail.ntua.gr}
\author{Georgios Varelogiannis}
\affiliation{Department of Physics, National Technical University
of Athens, GR-15780 Athens, Greece}

\vskip 1cm
\begin{abstract}

We report results from a systematic study of the competition of exotic triplet pair density wave (PDW) superconductivity (SC) with homogeneous (zero pair momentum) SC
in strongly polarized media such as half metallic systems. From the two
different PDW states allowed by symmetry in this
background only one may dominate
or even coexist with homogeneous SC.
We propose a direct experimental identification of PDW SC in this context. Our results suggest that these exotic states may plausibly emerge in
heterostructures involving proximity of SC with half-metallic CrO$_2$ 
where induced SC is established in the half metallic region and in strongly ferromagnetic superconductors.

\end{abstract}

\pacs{74.20.-z, 74.20.Rp}

\maketitle

The combination of ferromagnetism and superconductivity (SC) has been a fascinating challenge for decades \cite{Ginzburg,Maple}
since these states are expected to compete strongly.
The discovery of SC in the strongly ferromagnetic background of UGe$_2$ \cite{Saxena} has
revolutionized many of our ideas on the subject opening widely the field.
Additional FM superconductors have been discovered \cite{Aoki, Huy} throwing up challenging problems
such as reentrant SC \cite{Huxley} etc.
Equally challenging are the findings in heterostructures where proximity of SC and FM is enforced \cite{Kaizer,Wang, Khaire}.
  Indeed Kaizer \emph{et al}. reported a long distance supercurrent through the half metallic ferromagnet CrO$_2$ in contact with superconducting NbTiN
\cite{Kaizer}. Clearly, the proximity effect imposes bulk SC in CrO$_2$ despite its half metallic (fully magnetically polarized) character.

Understanding the type of exotic SC states that emerge in such extreme ferromagnetic conditions
is a challenge of great theoretical and practical importance.
Note that graphene nanoribbons are shown to be half metallic as well \cite{Son}
opening new avenues for the nanoengineering of such states.
It is natural to expect that at least in half metals SC is in the triplet channel where spins are parallel instead of being antiparallel as in usual singlet SC.
Up to now, only triplet SC states with zero pair momentum
have been considered in this context \cite{Efetov,Buzdin,Eschrig,Eschrig1,Volkov1}.
In the present Letter we argue that a different
type of triplet SC states, in which the pairs have finite momentum,
may dominate in the half metallic regions. These exotic SC states exhibit a density wave modulation of the superfluid density and we call them the triplet pair density wave (PDW) states or more formally the $\Pi$-triplet states.

The first studies of PDW SC in the singlet channel, also called $\eta$-pairing, were motivated by
its possible realization in the pseudogap regime of cuprates \cite{Zhang}. Recently,
experiments in the stripe-ordered materials La$_{1.875}$Ba$_{0.125}$CuO$_{4}$ 
and
La$_{1.6-x}$Nd$_0.4$Sr$_{x}$CuO$_{4}$ \cite{Li} renewed the interest for such states \cite{Berg}.
A PDW state in the triplet channel as the one considered here has been suggested to occur in the high field SC state of CeCoIn$_{5}$ coexisting with singlet SC and spin density waves \cite{Aperis} explaining
NMR \cite{Mitro} and fascinating neutron scattering results \cite{Kenzelmann}.
This state has also been considered in the same context by Sigrist and co-workers \cite{Yanase}.
Moreover PDW states have been suggested in the context of models for trapped fermionic gases \cite{Nikolic}.


We have examined systematically all the possible SC condensates that may emerge in a fully polarized medium and their competition within a microscopic mean field approach.
The Ginzburg-Landau theory for single-spin zero pair momentum SC has been done previously \cite{Pickett}.
Our starting point is a mean field BCS-type Hamiltonian:
\begin{align}
\label{hamiltonian}
\mathcal{H} &= \sum_{\mathbf{k}}\xi_{\mathbf{k}}\,c^{\dagger}_{\mathbf{k}}c_{\mathbf{k}}-
\sum_{\mathbf k} \bigl( \Delta^{\mathbf{0}}_{\mathbf{k}}\,c^{\dagger}_{\mathbf{k}}c^{\dagger}_{-\mathbf{k}} + \mathrm{h.c} \bigr) \notag \\
{}& \hspace{1.0cm}- \sum_{\mathbf k} \bigl(\Pi^{\mathbf{Q}}_{\mathbf k}\,c^{\dagger}_{\mathbf{k}}
c^{\dagger}_{-(\mathbf{k}+\mathbf{Q})}+ \mathrm{h.c} \bigr) 
\end{align}

\noindent The Hamiltionian (\ref{hamiltonian}) includes no spin since we assume that
we are in a strong ferromagnetic background thus all spins are frozen in the same direction.
The first term describes a tetragonal tight binding dispersion
which generically can be written as a sum of particle-hole symmetric terms and particle-hole asymmetric terms:
$\xi_{\bf k}=\gamma_{\bf k}+\delta_{\bf k}$.
When $\delta_{\mathbf{k}}=0$ there is particle-hole symmetry or perfect nesting with the \emph{commensurate}
wavevector $\mathbf{Q}$ while finite values of $\delta_{\mathbf{k}}$ destroy the nesting conditions.
The choice of a tetragonal dispersion is motivated by the fact that CrO$_2$ as well as strongly FM superconductors like UGe$_2$ and URhGe exhibit all a tetragonal structure.

The second term $\Delta^{\mathbf{0}}_{\mathbf{k}}= \sum_{\mathbf{k'}} V^{\mathbf{0}}_{\mathbf{k},\mathbf{k'}}
\left< c_{-\mathbf{k'}} c_{\mathbf{k'}} \right> $ represents unconventional SC with zero pair momentum,
and the last term  $\Pi^{\mathbf{Q}}_{\mathbf{k}}=\sum_{\mathbf{k'}} V^{\mathbf{Q}}_{\mathbf{k},\mathbf{k'}}
\left< c_{-(\mathbf{k'}+\mathbf{Q})} c_{\mathbf{k'}} \right> $ unconventional SC with finite pair momentum (PDW).
The effective interactions of the itinerant quasiparticles
$V^{\mathbf{0}}_{\mathbf{k},\mathbf{k'}}$,
$V^{\mathbf{Q}}_{\mathbf{k},\mathbf{k'}}$
may have a purely electronic origin in the case of FM superconductors.
In the case of heterostructures we assume within our approach that the effective potentials  incorporate the proximity effect as well. Naturally, we would expect in that case
a real space dependence of the potentials, that we neglect here. We only focus on qualitative symmetry questions that would not be affected by a smooth space dependence. Indeed the modulation of the superfluid density in our PDW SC state has a wavelength negligible compared to the coherence length or the characteristic lengths of the heterostructure.

To treat both SC
order parameters (OPs) in a compact
manner we introduce a Nambu-type representation using the spinors
$\Psi^{\dagger}_{\bf{k}}=\bigl(c^{\dagger}_{\bf{k}},
c_{-\bf{k}},c^{\dagger}_{\bf{k}+\bf{Q}},c_{-\bf{k}-\bf{Q}}\bigr) $.
Accordingly for the Nambu representation of the Hamiltonian
in Eq.~(\ref{hamiltonian}) we use the tensor products
$\widehat{\rho}_{i}=\bigl(\widehat{\sigma}_{i}\otimes \hat{1}_{2})~
\mbox{and}~ \widehat{\sigma}_{i}=\bigl(\hat{1}_{2} \otimes
\widehat{\sigma}_{i})$, where \(\widehat{\sigma}_{i}\) with \(i=1,2,3\) are the usual 2x2 Pauli matrices
and \( \hat{1}_{2} \) the unit 2x2 matrix. We classify the OPs with respect to their behavior under inversion $(\hat{I})$ $\mathbf{k} \rightarrow -\mathbf{k}$, translation
$(\hat{t}_{\bf {Q}})$ $\mathbf{k} \rightarrow \mathbf{k}+\mathbf{Q}$ and time reversal $(\hat{T})$.
Instead of the latter we may use complex conjugation $(\hat{K})$ which
is related to time reversal via the relations $\hat{T} \equiv -\hat{K}(\Delta^{\mathbf{0}}_{\mathbf{k}})$ and
$\hat{T} \equiv \hat{I} \hat{K}(\Delta^{\mathbf{Q}}_{\mathbf{k}})$.
Since the spins are frozen, the  $\mathbf{q}=0$ SC pair states may only have
odd parity: $\Delta^{\mathbf{0}}_{-\mathbf{k}} = -\Delta^{\mathbf{0}}_{\mathbf{k}}$.
Under translation we have both signs $\Delta^{\mathbf{0}}_{\mathbf{k+Q}}=
\pm\Delta^{\mathbf{0}}_{\mathbf{k}}$ and under $\hat{T}$ we get
$\hat{T} \Delta^{\mathbf{0}}_{\mathbf{k}}= - \Delta^{\mathbf{0}\,*}_{\mathbf{k}}$.
PDW states may have both parities $\Pi^{\mathbf{Q}}_{-\mathbf{k}}= \pm \Pi^{\mathbf{Q}}_{\mathbf{k}}$ and
both signs under translation since
$\Pi^{\mathbf{Q}}_{\mathbf{k+Q}}= -\Pi^{\mathbf{Q}}_{-\mathbf{k}} = \mp \Pi^{\mathbf{Q}}_{\mathbf{k}}$. Time reversal
demands that $\hat{T}\Pi^{\mathbf{Q}}_{\mathbf{k}}=\Pi^{\mathbf{Q}\,*}_{-\mathbf{k}}$ implying the relation $\hat{T}=\hat{I} \hat{K}$ for the
PDW states.

The time reversal symmetry is broken due to spin frozening.
This constrains us to consider only the states that are \emph{even in time reversal in our spinless formalism.}
The above symmetry properties allow {\it four possible SC
OPs, two with zero pair momentum $(\mathbf{q}=\mathbf{0})$ and two with finite pair momentum}:
$ \Delta^{\mathbf{0}I--}_{\mathbf{k}}, \hskip 0.3cm \Delta^{\mathbf{0}I-+}_{\mathbf{k}},
\hskip 0.3cm \Pi^{\mathbf{Q}I-+}_{\mathbf{k}}, \hskip 0.3cm
\Pi^{\mathbf{Q}R+-}_{\mathbf{k}}$.
Here the first index $\mathbf{q}=\mathbf{0}$ or $\mathbf{q}=\mathbf{Q}$ indicates the {\it total
momentum of the pair}, the second index $R$ or $I$ indicates whether
the OP is real or imaginary, the third index $\pm$
indicates parity under inversion $\hat{I}$ and the last index denotes gap
symmetry under $\hat{t}_{\mathbf{Q}}$.
The symmetry properties of the OPs under inversion \(\hat{I} \)
and translation \(\hat{t}_{\mathbf{Q}} \) imply a specific structure in \(\mathbf{k}\)-space.
Every OP $M_{\mathbf{k}}$ is written in the form $M_{\mathbf{k}}=M f_{\mathbf{k}}$
where the form factors $f_{\mathbf{k}}$ belong to different irreducible
representations of the tetragonal group D$_{4h}$. Specifically:  $\Delta^{\mathbf{0}I--}_{\mathbf{k}} \sim \sin k_{x} + \sin k_{y}$ (s-wave),
$\Delta^{\mathbf{0}I-+}_{\mathbf{k}}, \Pi^{\mathbf{Q}I-+}_{\mathbf{k}} \sim sin (k_{x}+k_{y})$ (p-wave) and
$\Pi^{\mathbf{Q}R+-}_{\mathbf{k}} \sim \cos k_{x} - \cos k_{y}$ (d-wave).

According to the above symmetry classification there exist
four possible pairs of competing SC states with zero and finite pair momentum.
Using our formalism we calculate Green's functions and self-consistent systems of
gap equations for each case. The competition of $ \Delta^{\mathbf{0}I--}_{\mathbf{k}}$ with $\Pi^{\mathbf{Q}R+-}_{\mathbf{k}} $ and
$ \Delta^{\mathbf{0}I--}_{\mathbf{k}}$ with $\Pi^{\mathbf{Q}I-+}_{\mathbf{k}} $
obey the following system of equations:

\begin{align}
\label{zones1ZCgap} \Delta_{\mathbf{k}} &= \sum_{\mathbf{k'}} V^{\Delta}_{\mathbf{k},\mathbf{k'}} \Delta_{\mathbf{k'}}
\Biggl\{ \frac{1}{4E_{+}(\mathbf{k'})} \tanh \biggl({ E_{+}(\mathbf{k'})\over 2T }\biggr) \notag \\
&{} \hspace{2.0cm} + \frac{1}{4E_{-}(\mathbf{k'})}
\tanh \biggl( { E_{-}(\mathbf{k'})\over 2T}\biggr) \Biggr\}\\
\label{zones1ZBgap} \Pi_{\mathbf{k}} &= \sum_{\mathbf{k}'} V^{\Pi}_{\mathbf{k}, \mathbf{k'}} \Pi_{\mathbf{k}'}
\Biggl\{ {A(\mathbf{k}') + \gamma_{\mathbf{k}'}
\over 4 E_{+}({\mathbf{k}'})A(\mathbf{k}')}
\tanh \biggl( {E_{+} ({\mathbf{k}'})\over 2T} \biggr) \nonumber\\
&{} \hspace{1.0cm} + {A(\mathbf{k}') - \gamma_{\mathbf{k}'}
\over 4 E_{-}({\mathbf{k}'})A(\mathbf{k}')}
\tanh \biggl( {E_{-} ({\mathbf{k}'})\over 2T} \biggr)\Biggr\}
\end{align}

\noindent where $A(\mathbf{k}) \equiv \sqrt{\delta_{\mathbf{k}}^2 + \Pi_{\mathbf{k}}^2}$ and the quasiparticle energies are given by:

\begin{equation}
\label{zones1pole}
E_{\pm}(\mathbf{k}) = \sqrt{\Bigl(\sqrt{\delta^2_{\mathbf{k}}+ \Pi^2_{\mathbf{k}}} \pm \gamma_{\mathbf{k}} \Bigr)^{2}
+\Delta^2_{\mathbf{k}}}
\end{equation} \\

\noindent  The remaining two cases, competition of $ \Delta^{\mathbf{0}I-+}_{\mathbf{k}}$ with $\Pi^{\mathbf{Q}R+-}_{\mathbf{k}} $ and
$ \Delta^{\mathbf{0}I-+}_{\mathbf{k}}$ with $\Pi^{\mathbf{Q}I-+}_{\mathbf{k}} $, obey the following equations:

\begin{align}
\label{zones2ZCgap}
&\Delta_{\mathbf{k}} = \sum_{\mathbf{k'}} V^{\Delta}_{\mathbf{k},\mathbf{k'}} \Delta_{\mathbf{k'}}
\Biggl\{ \frac{B(\mathbf{k}')+ \Pi^{2}_{\mathbf{k'}}}{4E_{+}(\mathbf{k'})B(\mathbf{k}')}
\tanh \biggl({ E_{+}(\mathbf{k'})\over 2T }\biggr) \nonumber \\
&{} \hspace{1.5cm} + \frac{B(\mathbf{k}')- \Pi^{2}_{\mathbf{k'}}}{4E_{-}(\mathbf{k'})B(\mathbf{k}')}
\tanh \biggl({ E_{-}(\mathbf{k'})\over 2T }\biggr) \Biggr\}  \\
\label{zones2ZBgap}
&\Pi_{\mathbf{k}} = \sum_{\mathbf{k}'} V_{\mathbf{k}, \mathbf{k'}}^{\Pi} \Pi_{\mathbf{k}'}
\Biggl\{ \frac{B(\mathbf{k}')+ \gamma^{2}_{\mathbf{k'}}+ \Delta^{2}_{\mathbf{k'}}}{4E_{+}(\mathbf{k'})B(\mathbf{k}')} \tanh \biggl({ E_{+}(\mathbf{k'})\over 2T }\biggr) \nonumber \\
&{} \hspace{1.5cm} + \frac{B(\mathbf{k}')- \gamma^{2}_{\mathbf{k'}}- \Delta^{2}_{\mathbf{k'}}}{4E_{+}(\mathbf{k'})B(\mathbf{k}')} \tanh \biggl({ E_{-}(\mathbf{k'})\over 2T }\biggr)\Biggr\}
\end{align}

\noindent where $B(\mathbf{k}) \equiv \sqrt{\gamma^{2}_{\mathbf{k}}(\delta^{2}_{\mathbf{k}} + \Pi^{2}_{\mathbf{k}}) + \Delta^{2}_{\mathbf{k}}\Pi^{2}_{\mathbf{k}}}$ and the dispersion relations take the form:

\begin{equation}
\label{zones2pole}
E_{\pm}({\bf k}) = \sqrt{ { \Delta_{\mathbf{k}}^2 \delta_{\mathbf{k}}^2 \over
\delta_{\mathbf{k}}^2 + \Pi_{\mathbf{k}}^2}+ \Biggl( \sqrt{ \delta_{\mathbf{k}}^2 +
\Pi_{\mathbf{k}}^2}\pm \sqrt{\gamma_{\mathbf{k}}^2 + { \Delta_{\mathbf{k}}^2
\Pi_{\mathbf{k}}^2\over \delta_{\mathbf{k}}^2 + \Pi_{\mathbf{k}}^2}}\,\Biggr)^2}
\end{equation} \\

\noindent The effective potentials $V^{\Delta}_{\mathbf{k},\mathbf{k'}}, V^{\Pi}_{\mathbf{k},\mathbf{k'}}$ have the form
$V_{\mathbf{k},\mathbf{k'}}= V f_{\mathbf{k}}f_{\mathbf{k}'} $ (separable potentials).
We have solved self consistently the systems of equations (\ref{zones1ZCgap}), (\ref{zones1ZBgap})
and (\ref{zones2ZCgap}), (\ref{zones2ZBgap}) on a square lattice with $\gamma_{\mathbf{k}}= -t_1 (\cos k_x + \cos k_y)$ and
$\delta_{\mathbf{k}}=- t_2\cos k_x \cos k_y$ and $\mathbf{Q}=(\pi,\pi)$.
For every competing pair we have performed a large number of self-consistent calculations
varying the pairing potential in the two channels, the temperature
and the ratio $t_{2}/t_{1}$.

\begin{figure}[!h]
\includegraphics[width=7.5cm,height=10.0cm,angle=0]{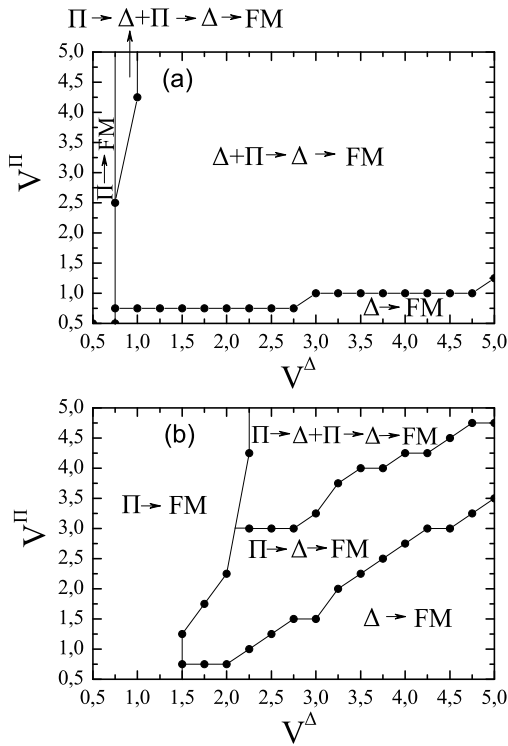}
\caption{Maps of the dependence of phase sequences on the effective interactions
$V^{\Delta}$ and $V^{\Pi}$ for low temperature. Arrows indicate
the cascade of phases obtained when $t_2/t_1$ grows starting from zero. The black dots
separate regions of different phase {\it sequences} under growing $t_2/t_1$.
All phases coexist with ferromagnetism (FM). The phases
indicated as FM, are phases in which there is not any finite $\Delta$ or
$\Pi$ OP and so only FM is present. Panel (a) corresponds
to the competition of $\Pi^{\mathbf{Q}R+-}$ with $\Delta^{\mathbf{0}I--}$. Panel (b)
corresponds to the competition of $\Pi^{\mathbf{Q}R+-}$ with $\Delta^{\mathbf{0}I-+}$.
The potentials are in units of $t_1$.} \label{fig:maps}
\end{figure}

The first important result is that the $\Pi^{\mathbf{Q}I-+}_{\mathbf{k}}$ OP
can never survive when it competes with any of the two zero pair momentum SC states.
Specifically, the $\Pi^{\mathbf{Q}I-+}_{\mathbf{k}}$ gap is zero regardless of the values of the pairing potentials and the
particle-hole asymmetry term.
Therefore, \emph{although the state $\Pi^{\mathbf{Q}I-+}_{\mathbf{k}}$ is allowed by symmetry it is never realized}.
We report here results from the competition of the remaining PDW OP $\Pi^{\mathbf{Q}R+-}_{\mathbf{k}} $ with both zero momentum SC states.
The phase sequences as $t_{2}/t_{1}$ grows starting from zero with respect to the values of the effective potentials for the
competition $\Pi^{\mathbf{Q}R+-}_{\mathbf{k}} $ with $\Delta^{\mathbf{0}I--}_{\mathbf{k}}$ and
$\Pi^{\mathbf{Q}R+-}_{\mathbf{k}} $ with $\Delta^{\mathbf{0}I-+}_{\mathbf{k}}$ are shown in Fig. \ref{fig:maps}.
Arrows in Fig. \ref{fig:maps} indicate the cascade of phases
observed when the ratio $t_{2}/t_{1}$ grows starting from zero.
Since we consider a spin-polarized background, all states reported also coexist with FM,
and the transitions to the FM state reported at high values of $t_{2}/t_{1}$ has the meaning
of a transition to a state that is only ferromagnetic with
no SC OP present. \\

The competition of $\Pi^{\mathbf{Q}R+-}_{\mathbf{k}}$ with $\Delta^{\mathbf{0}I--}_{\mathbf{k}}$ favors the \emph{coexistence} of both
SC states $(\mathbf{q}=\mathbf{0} \; \text{and} \; \mathbf{q}=\mathbf{Q})$ in low-T
over a wide range of values of the pairing potentials (Fig. \ref{fig:maps}a).
The transition from a coexistence state to a SC state with $\mathbf{q}=\mathbf{0}$
 when $t_2/t_1$ grows is always continuous (\emph{second order}) and dominates the $V^{\Delta},V^{\Pi}$ parameter space.

\begin{figure}[!h]
\centerline{\includegraphics[width=7cm,height=8cm]{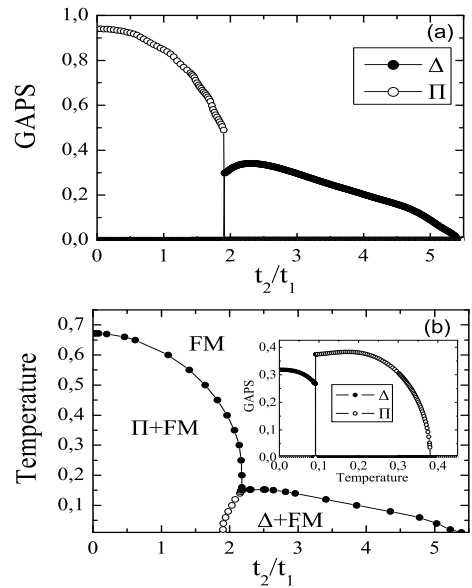}}
\caption {{\footnotesize (a) Dependence of $\Delta^{\mathbf{0}I-+}_{\mathbf{k}}$
and $\Pi^{\mathbf{Q}R+-}_{\mathbf{k}} $
on \(t_{2}/t_{1}\) in low-T. (b) $t_{2}/t_{1}$-temperature phase diagram.
Closed symbols mark 2nd order and open symbols 1st order transitions.
A transition \emph{within the SC phase}, from the PDW state to the $\Delta^{\mathbf{0}I-+}_{\mathbf{k}}$ state,
is possible with decreasing temperature.
The values of the pairing potentials are  $V^{\Delta}=V^{\Pi}=3$.  }} \label{ZCI-+ZBR+-:phase}
\end{figure}

The low temperature regime is different when $\Pi^{\mathbf{Q}R+-}$ competes with $\Delta^{\mathbf{0}I-+}$.
Coexistence of the two SC states is allowed again but is restricted only to a small portion of $V^{\Delta},V^{\Pi}$
space (Fig. \ref{fig:maps}b) compared with the previous case.
The most interesting feature of the $V^{\Delta},V^{\Pi}$ map is \emph{the domination
of the PDW state for the smaller values of $t_2/t_1$.}. Thus, in this case the formation of
the $\Pi^{\mathbf{Q}R+-}$ PDW state is favored. As particle hole asymmetry grows ($t_{2}/t_{1}$ grows) we may have transitions from PDW to a state of coexistence or to a zero pair momentum SC
state.
We report in Fig. \ref{ZCI-+ZBR+-:phase} the dependence of the OPs on $t_{2}/t_{1}$
 at low-T (Fig. \ref{ZCI-+ZBR+-:phase}a) and the phase diagram (Fig. \ref{ZCI-+ZBR+-:phase}b) obtained for $V^{\Delta}=V^{\Pi}=3$.
These values correspond to the transition $\Pi \rightarrow \Delta \rightarrow FM$ of Fig. \ref{fig:maps}b.
 At low-T the transition from the PDW to the $\Delta$ state is \emph{first order} in $t_{2}/t_{1}$, and we note that the PDW gap is significantly larger than the
$\Delta$ gap although the pairing potentials have the same magnitude (Fig. \ref{ZCI-+ZBR+-:phase}a).
The phase diagram shows that the transition $\Pi \rightarrow \Delta$ with $t_{2}/t_{1}$ is not limited to low-T.
The PDW phase extends to higher temperatures (Fig. \ref{ZCI-+ZBR+-:phase}b) than the $\Delta$-phase.
The boundary separating the two SC states is \emph{first order} and ends at a tricritical point.
Decreasing the temperature moves the boundary to lower $t_{2}/t_{1}$-values.
This allows a \emph{first order}  transition with respect to temperature \emph{within the superconducting phase} from the $\Pi$ to $\Delta$ state.
An example of such a transition realized for $t_{2}/t_{1}=2$ is shown in the inset of Fig. \ref{ZCI-+ZBR+-:phase}b. 

\begin{figure}[!h]
\includegraphics[width=4cm,height=4cm]{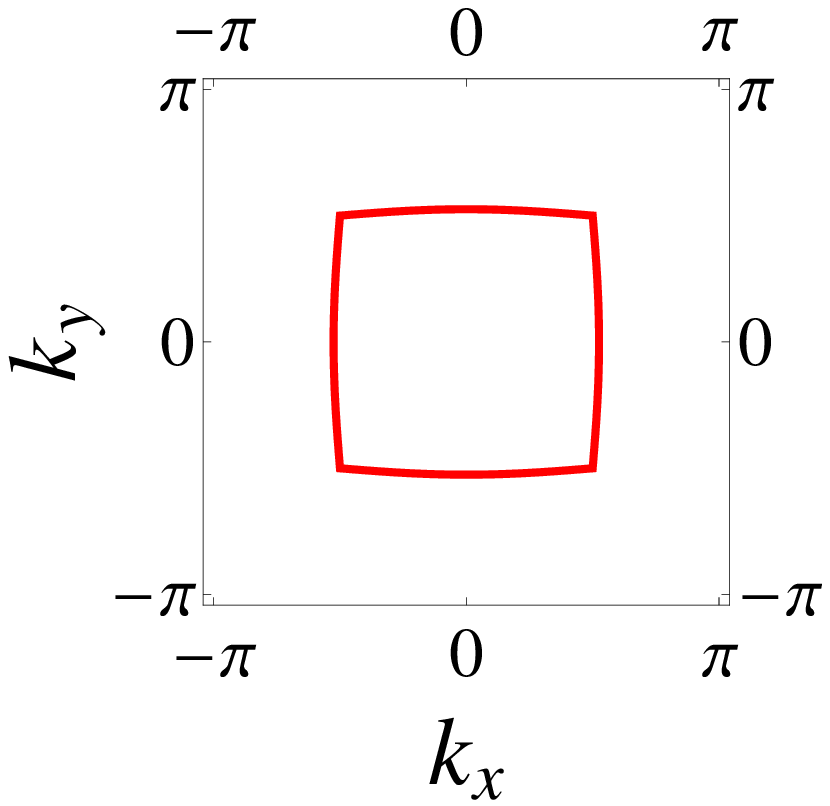}
\includegraphics[width=4cm,height=4cm]{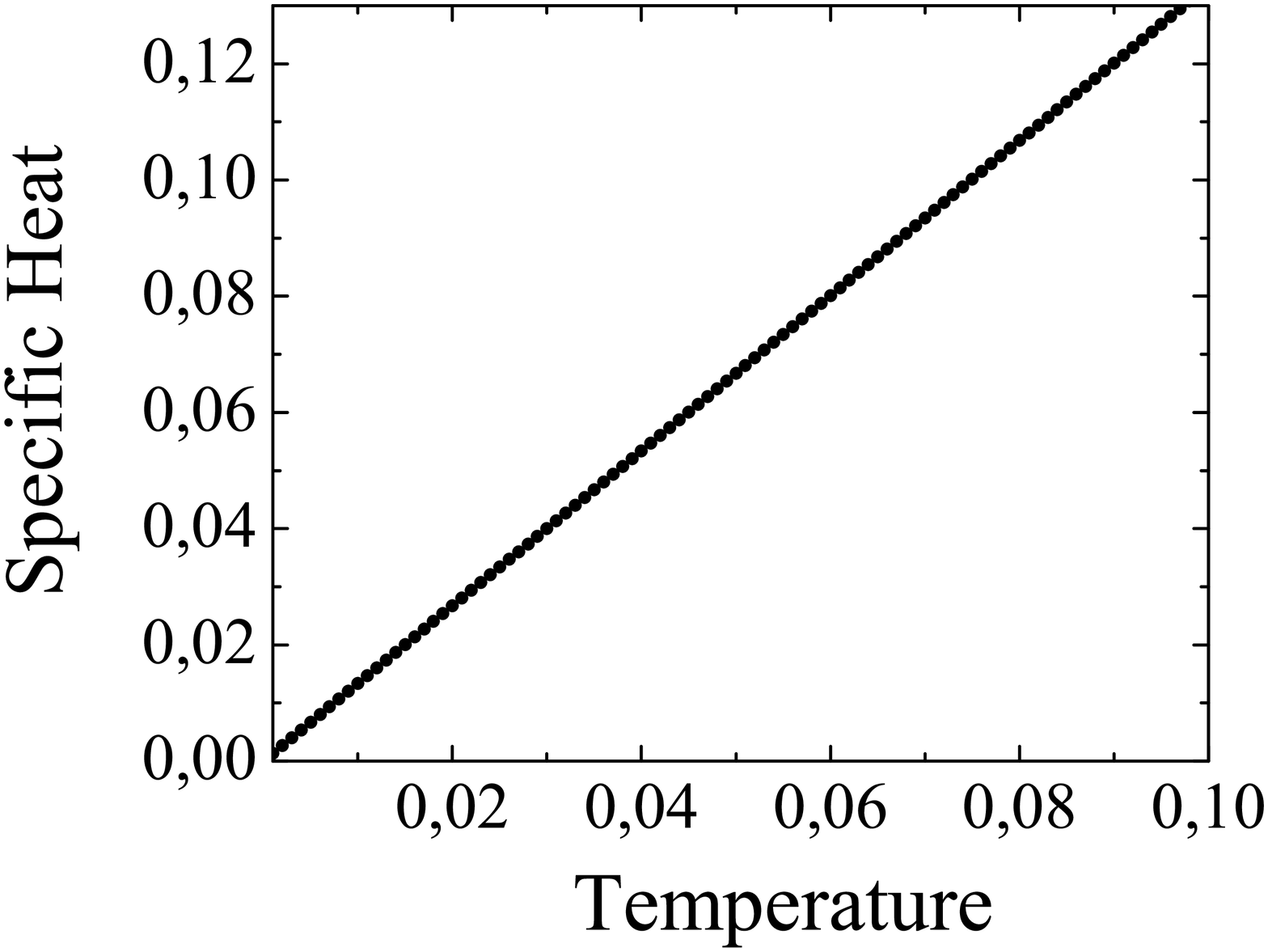}
\includegraphics[width=4cm,height=4cm]{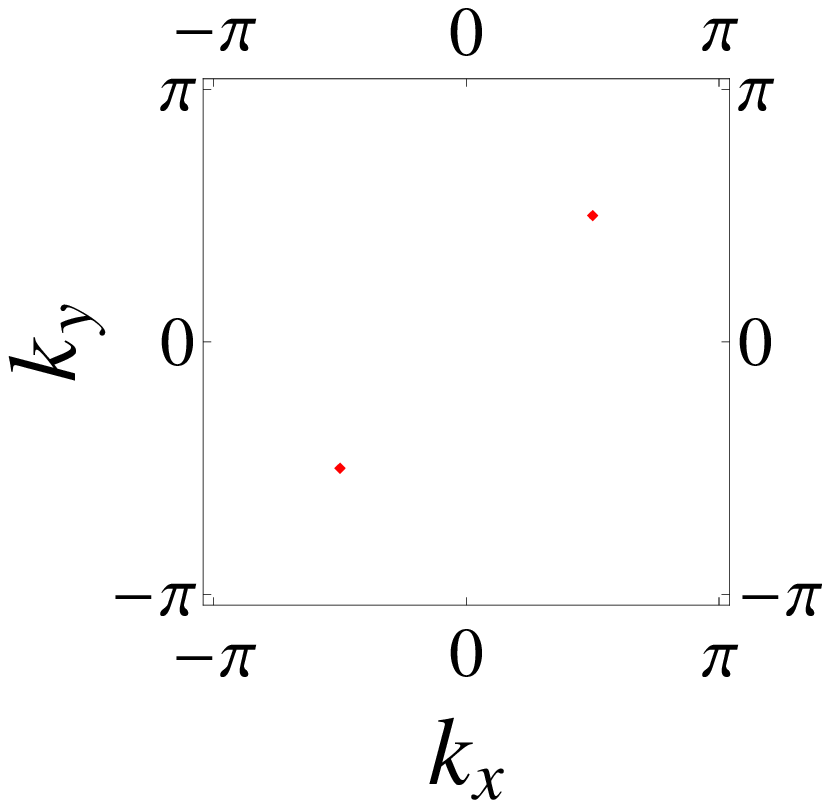}
\includegraphics[width=4cm,height=4cm]{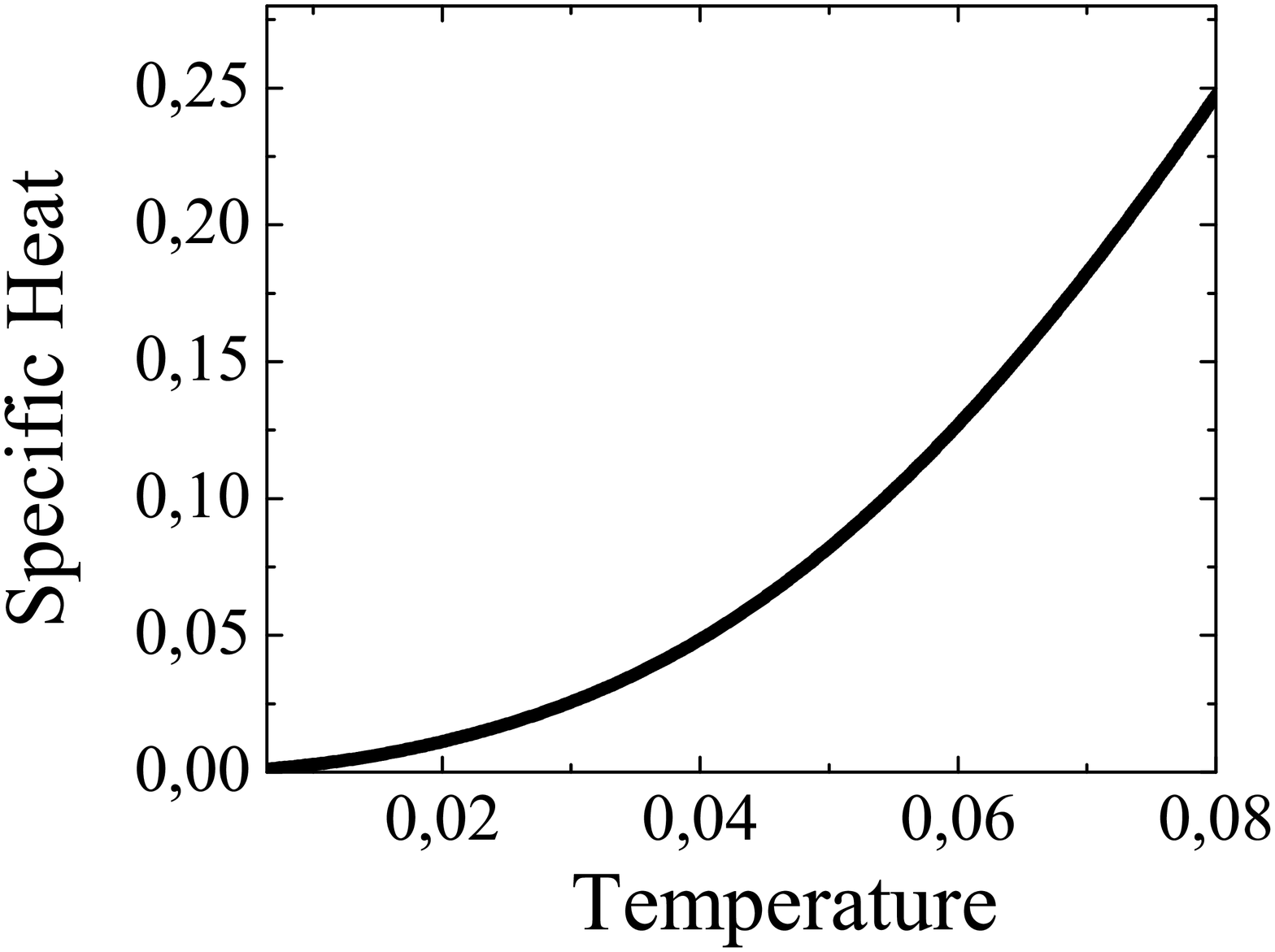}
\caption{{\footnotesize Fermi surface (left) and specific heat at low-T  (right) in the $\Pi^{\mathbf{Q}R+-}_{\mathbf{k}}$ phase  $t_{2}/t_{1}=1.0$ (upper panel)
and in the $\Delta^{\mathbf{0}I-+}_{\mathbf{k}}$-phase $t_{2}/t_{1}=2.5$ (lower panel).
} } \label{fig:FermiCv}
\end{figure}

A question that naturally emerges is how the exotic PDW state $\Pi^{\mathbf{Q}R+-}_{\mathbf{k}}$
can be identified experimentally.
Quite remarkably, specific heat measurements at low-T may be enough.
Both SC states of zero pair momentum exhibit a polynomial behavior of the specific heat in the low temperature regime and this also the case when $\Pi^{\mathbf{Q}R+-}_{\mathbf{k}}$ coexists with
either of the two SC states of zero pair momentum. On the other hand when purely
$\Pi^{\mathbf{Q}R+-}_{\mathbf{k}}$ state is present
the specific heat at low-T exhibits a linear temperature behavior.
We illustrate that in Fig. \ref{fig:FermiCv} where the Fermi surface and the specific heat
for $t_{2}/t_{1}=1.0$  in the PDW-phase (upper panel) and
$t_{2}/t_{1}=2.5$  in the $\Delta$-phase (lower panel) are reported. We observe that in
the $\Pi^{\mathbf{Q}R+-}_{\mathbf{k}}$ phase the
Fermi surface is extended imposing a linear specific heat at low-T.
On the contrary, in the
$\Delta^{\mathbf{0}I-+}_{\mathbf{k}}$ phase we have only two Fermi points and the specific heat at low-T exhibits a polynomial dependence. This is also the case for the other homogeneous SC state $\Delta^{\mathbf{0}I--}_{\mathbf{k}}$. Therefore \emph{linear low-T specific heat identifies the PDW state}.

In summary, we explored systematically the possibility that
exotic triplet PDW states
may dominate or coexist with usual triplet SC states of zero pair momentum in a fully polarized electronic system.
We find that two SC states with zero ($\mathbf{q}=\mathbf{0}$)
and two SC states with finite ($\mathbf{q}=\mathbf{Q}$) pair momentum
are allowed by symmetry in this context.
Our calculations showed that the PDW state $\Pi^{\mathbf{Q}I-+}_{\mathbf{k}}$
having the p-wave symmetry can never survive when it competes with any
of the two SC states of zero pair momentum.
The other PDW $\Pi^{\mathbf{Q}R+-}_{\mathbf{k}}$ having
d-wave symmetry may \emph{either appear alone dominating upon the SC states of zero pair momentum or coexist with them}.
Specific heat measurements in low-T can identify this exotic PDW phase that may plausibly develop in Superconductor - half metal heterostructures and in strongly ferromagnetic superconductors as well.

We are grateful to A. Aperis, P. Kotetes and P. Thalmeier for illuminating discussions.

\end{document}